\documentclass[aps,prl,twocolumn,showpacs,floatfix]{revtex4-1}
\usepackage{graphicx}
\usepackage{amssymb}
\usepackage{dcolumn}
\usepackage{bm}
\usepackage{dcolumn}
\usepackage{url}

\def\s#1{_{\rm #1} }

\def\lz{\ell_{\parallel}}
\def\lp{\ell_{\bot}}
\def\Ppa{P_{\parallel}}
\def\Ppr{P_{\bot}}
\def\bea{\begin{eqnarray}}
\def\eea{\end{eqnarray}}
\def \be{\begin{equation}}
\def \ee{\end{equation}}

\begin{document}
\title{Optomechanical elastomeric engine}
\author{Milo\v{s} Kne\v{z}evi\'{c}}
\email{mk684@cam.ac.uk}
\author{Mark Warner}
%\email{mw141@cam.ac.uk}
\affiliation{Cavendish Laboratory, University of Cambridge, Cambridge CB3 0HE, United Kingdom}
\date{\today}

\begin{abstract}
Nematic elastomers contract along their director when heated or illuminated (in the case of photoelastomers).
We present a conceptual design for an elastomer-based engine to extract mechanical work from heat or light.
The material parameters and the geometry of such an engine are explored, and it is shown that its efficiency
can go up to 20\%.
\end{abstract}


\maketitle

Efficiently converting solar energy to mechanical or electrical energy is one of
the greatest contemporary challenges in science and technology. In this Letter we propose an engine
based on liquid crystal elastomers (LCEs)~\cite{warnerbook:07} that extracts mechanical work from heat or light.
As first intimated by de Gennes~\cite{deGennes:75}, unusual properties of LCEs arise from a coupling between
the liquid crystalline ordering of mesogenic molecules and the elasticity of the underlying
polymer network. Various external stimuli, in particular heat or light cause reversible contractions of monodomain LCEs
along their nematic director, with recovery elongations on stimuli removal. The shape changes of the sample can be remarkable -- up
to 350\% and occur in a relatively narrow temperature interval around the nematic-isotropic transition
temperature~\cite{finkelmann:00,tajbakhsh:01}.  The contraction-elongation cycle can be repeated many times, and be
exploited to construct a continuously operating engine in which heat or light is used to produce mechanical work.

Cross-linked networks of polymer chains of a LCE include mesogenic units that belong to either the polymer backbone
(main-chain LCE) or side units pendent to the backbone (side-chain LCE)~\cite{warnerbook:07}. The shape of a monodomain
nematic LCE strongly depends on the temperature-dependent nematic order parameter $Q(T)$, due to the coupling of $Q$ with the
average polymer chain anisotropy. Increasing the temperature decreases $Q$, causing a decrease of the
polymer backbone anisotropy, which manifests as a uniaxial contraction of the sample.

Mechanical change of a LCE can also be achieved by introducing photoisomerizable dye molecules into its
chemical structure (nematic photoelastomers~\cite{finkelmann:01,hogan:02}). Upon illumination, dye molecules can
undergo transitions from their linear (\textit{trans}) ground state to the excited bent-shaped (\textit{cis}) state.
The rodlike \textit{trans} molecules contribute to the overall nematic order, while the bent \textit{cis} molecules
act as impurities that reduce the nematic order parameter, in turn leading to a macroscopic contraction.

\begin{figure}[b]
  \includegraphics[width=8.7cm]{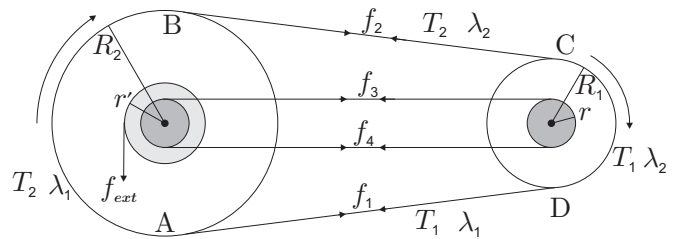}
  \caption{Schematic of an optomechanical LCE engine.
  }
\label{fig1}
\end{figure}

The operating principle of an LCE engine is shown in Fig.~\ref{fig1}. A closed band of nematic elastomer of initial
length $L_0$ is stretched and wound around two pulleys of radii $R_1$ and $R_2$ ($R_2> R_1$).
Initially, the whole elastomeric band is in the nematic state at some temperature $T_1$. The transmission pulleys
of equal radii $r$, rigidly coupled with the main wheels, are connected by a loop of inextensible string.
Obviously, if the temperature of the whole system is $T_1$, in the absence of external forces, the system is at
rest. By increasing the temperature of a part of the elastomer to a value $T_2$, an excess contractile force, $f_2 - f_1$,
will occur (see Fig.~\ref{fig1}). This force acts on wheels of radii $R_1$ and $R_2$ and tends to rotate the former counter-clockwise
and the latter in a clockwise direction; since $R_2>R_1$ the wheels will turn clockwise.
The rotation brings a piece of elastomer initially being at temperature $T_1$ to the temperature $T_2$,
while an another piece of elastomer having temperature $T_2$ returns to the temperature $T_1$. By keeping the temperatures
$T_1$ and $T_2$ at fixed values, this process can be reproduced many times, which provides the basis for a continuous
operation of the engine. The engine operation cycle is reminiscent of an engine based on chemomechanical conversion~\cite{steinberg:66}.
Our stretch engine is quite different from LCE bend motors~\cite{ikeda:08,palffy:13}.

Mechanical work can be obtained by applying a suitable external force $f\s{ext}$, for example by attaching a
weight to the end of a thread wound around a pulley of radius $r'$ (see Fig.~\ref{fig1}). During the engine operation,
a part of the energy invested to heat the elastomer to the temperature $T_2$ is converted into mechanical work.
An another way to realize such an engine is based on the use of photoelastomers. In this case illumination causes the
creation of \textit{cis} isomers, which in turn can be seen as a light-dependent increase of the actual temperature
$T_1$ to the new, now effective, value $T_2$~\cite{finkelmann:01}. Our analysis applies to both thermo-
and photo-engines.

We shall assume that elastomer coming in contact with the wheel of radius $R_2$ changes its temperature from
$T_1$ to $T_2$ before leaving the wheel, and stays at $T_2$ until it hits the wheel $R_1$. The engine in Fig.~\ref{fig1}
requires heating in the part AB (illumination in the case of photoelastomers),
while in CD cooling to $T_1$ should be ensured (relaxation to the dark state). Parts BC and DA should be also kept at the
constant temperatures $T_2$ and $T_1$, respectively.

In steady regime, the amount of elastomer taken on to the wheel of radius $R_2$ should equal
the amount taken on to the wheel of radius $R_1$, that is $\Delta \theta R_2 \rho/ \lambda_1 = \Delta \theta R_1 \rho/ \lambda_2$.
Here, $\rho$ is the linear density of elastomer in the formation state at temperature $T_1$, $\Delta \theta$ denotes the rotation angle of the wheels, and $\lambda_1$ and $\lambda_2$ are the stretches in the parts DA and BC of the engine, respectively (stretches are measured from
the formation state). We assume that elastomer in contact with the wheel does not slip, and does not change its length even if it
experiences change of temperature, i.e, stretch remains equal to $\lambda_1$ in the part AB, and equal to $\lambda_2$ in the part CD.
The above condition can be rewritten as
\be
\gamma = \frac{R_2}{R_1} = \frac{\lambda_1}{\lambda_2} \ge 1,
\label{ratio}
\ee
where the ratio of the wheel radii is denoted by $\gamma$.

To avoid slack, the length $L_0$ of the elastomer loop in the formation state should be smaller than approximately
$2L + \pi (R_1 + R_2)$, where $L$ is the distance BC. Since there is the stretch $\lambda_1$ in the part
DAB of the elastomer and $\lambda_2$ in its remaining part BCD, one can write
\be
\frac{1}{\lambda_1} (L+ \pi R_2) + \frac{1}{\lambda_2} (L + \pi R_1) = L_0.
\label{geometry}
\ee
Relations (\ref{ratio}) and (\ref{geometry}) allow one to express the stretch $\lambda_2$ via
reduced lengths $\widetilde{L} = L/\pi R_1$ and $\widetilde{L}_0 = L_0/\pi R_1$,
\be
\lambda_2 = \frac{1}{\widetilde{L}_0} \left [\widetilde{L} \left (1 + \frac{1}{\gamma} \right ) + 2 \right ].
\label{l2}
\ee

The inextensible inner wire on wheels of radii $r$ forces the angular velocities of wheels to be equal (see Fig.~\ref{fig1}).
When the engine runs at a constant velocity the net torque acting on each of the pulleys is zero.
Neglecting frictional forces at the bearings, the balance of torques on wheels of radii $R_1$ and $R_2$ is respectively:
\bea
(f_2 - f_1)R_1 + (f_3 - f_4)r &=& 0, \nonumber \\
(f_1 - f_2)R_2 + (f_4 - f_3)r + f\s{ext} r' &=& 0,
\label{torques}
\eea
where $f_3$ and $f_4$ are the forces acting on the wheels of radii $r$ (Fig.~\ref{fig1}). From these two equations we get
\be
G\s{ext} \equiv f\s{ext} r' = (f_2 - f_1)(R_2 - R_1),
\label{gext}
\ee
where $G\s{ext}$ is the magnitude of the torque of the external force $f\s{ext}$. In what follows we express the forces
$f_1$ and $f_2$ in terms of stretches $\lambda_1$ and $\lambda_2$.

Due to the presence of mesogenic molecules, long polymer chains of nematic elastomers have an anisotropic Gaussian distribution.
The elastic free energy density of a nematic rubber in response to a deformation $\lambda$ along
the director $\underline{n}$ can be written in the form~\cite{warnerbook:07}
\be
F = \frac{1}{2} \mu \left ( \lambda^2 \frac{\lz^{(1)}}{\lz^{(2)}} + \frac{2}{\lambda} \frac{\lp^{(1)}}{\lp^{(2)}} \right ),
\label{fenergy}
\ee
where $\mu$ is the shear modulus in the isotropic state.
The Flory step lengths in directions parallel and perpendicular to the director $\underline{n}$
have different values $\lz$ and $\lp$ (the director is along the long direction of the elastomeric band).
We assume that the elastomer is formed at $T_1$ (corresponding step lengths
are $\lz^{(1)}$ and $\lp^{(1)}$), and has current step lengths $\lz^{(2)}$ and $\lp^{(2)}$ (for example, at $T_2$).
Given that we are concerned only with derivatives of $F$ with respect to $\lambda$, we omitted $\lambda$-independent terms in Eq.~(\ref{fenergy}).
As rubber changes shape at constant volume, the area of the elastomer perpendicular to the director $\underline{n}$
changes by a factor of $1/\lambda$.

The force exerted by an elastomer is proportional to the derivative of free energy density with respect to stretch,
$f = A_0 (\partial F/\partial \lambda)_T$, where $A_0$ is the area of the cross-section of the elastomer in the formation state.
For the part DA of elastomer one has $\lambda = \lambda_1$, $\lz^{(2)}=\lz^{(1)}$ and $\lp^{(2)}=\lp^{(1)}$, while for the part BC
one has $\lambda = \lambda_2$, with $\lz^{(2)}$ and $\lp^{(2)}$ taking values smaller and larger than
$\lz^{(1)}$ and $\lp^{(1)}$ respectively (for prolate symmetry elastomers). Then the forces $f_1$ and $f_2$ are
\bea
f_1 = \mu(T_1) A_0 \left ( \lambda_1 - \frac{1}{\lambda_1^2} \right ), \nonumber \\
f_2 = \mu(T_2) A_0 \left ( \lambda_2 \Ppa - \frac{\Ppr}{\lambda_2^2} \right ),
\label{forces}
\eea
where the ratios $\Ppa = \lz^{(1)}/\lz^{(2)}>1$ and $\Ppr = \lp^{(1)}/\lp^{(2)}<1$ depend on the order parameters $Q_1$ and $Q_2$.
Note that a free elastomer heated from temperature $T_1$ to $T_2$ undergoes the natural contraction
$\lambda\s{m} = (\Ppr/\Ppa)^{1/3}$ along its director (this relation can be obtained by setting $f_2=0$ in the above equation).

The isotropic moduli appearing in equations (\ref{forces}) are assumed to be comparable, $\mu(T_1) \approx \mu(T_2)$.
On inserting (\ref{forces}) into (\ref{gext}), the reduced torque $\mathcal{G} = (f_2 - f_1)(R_2 - R_1)/ \mu A_0 R_1$ is
\be
\mathcal{G} = (\gamma - 1) \left [ \lambda_2 (\Ppa - \gamma) - \frac{1}{\lambda_2^2} \left ( \Ppr - \frac{1}{\gamma^2} \right ) \right ].
\label{reducedg}
\ee

\begin{figure}[b]
  \includegraphics[width=6.7cm]{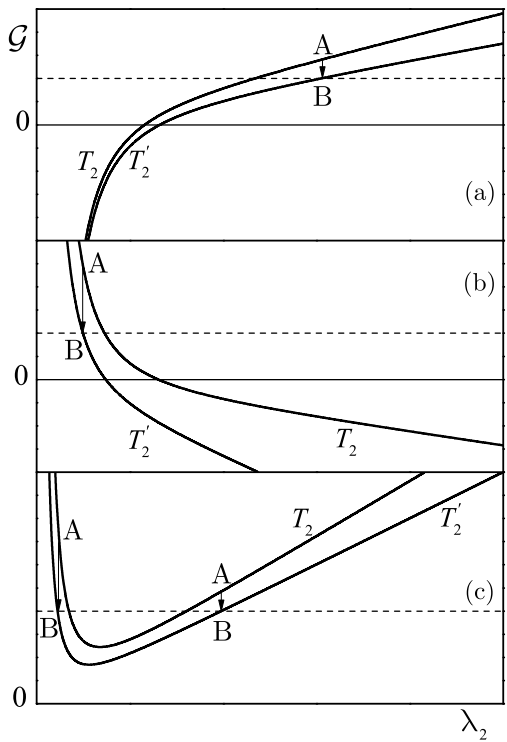}
  \caption{Typical dependence of the reduced torque $\mathcal{G}$ on stretch $\lambda_2$ corresponding to
  cases (a) -- (c) in the main text. The reduced external torques $G\s{ext}/\mu A_0 R_1$ are represented by the dashed lines.
  }
\label{fig2}
\end{figure}

We compare this torque to that of the reduced torque $G\s{ext}/\mu A_0 R_1$ from the external forces.
Since the ratio $\gamma$ of the wheel radii is greater than 1, then $\mathcal{G}>0$ if $[ \dots ]$ of (\ref{reducedg}) is positive.
We examine four different cases: \\
(a) $\Ppa - \gamma > 0$ and $\Ppr - 1/\gamma^2 > 0$, which is equivalent to $\Ppa > \gamma > 1/\sqrt{\Ppr}$,
involving a purely material condition $\Ppa \sqrt{\Ppr} >1$.
The reduced torque $\mathcal{G}$ as a function of $\lambda_2$ is shown in Fig.~\ref{fig2}(a) for two different temperatures $T_2$ and
$T_2'$ ($T_2'<T_2$). It is easy to see that $\mathcal{G}$ vanishes for $\lambda_2 = [(\Ppr - 1/\gamma^2)/(\Ppa - \gamma)]^{1/3}$.
At the point A the torque $\mathcal{G}$ is greater than the reduced torque of external force $G\s{ext}/\mu A_0 R_1$, and
the engine turns more quickly until it does not have time to heat to the temperature $T_2$. It only gets to temperature
$T_2' < T_2$ and moves on to $T_2'$ curve at the point B.
This governing of the delivered torque by speed of rotation is reminiscent of an electric motor; rotation-induced
back electromotive force limits current flow and hence limits torque.\\
(b) $\Ppa - \gamma < 0$ and $\Ppr - 1/\gamma^2 < 0$, which can be expressed as $\Ppa < \gamma < 1/\sqrt{\Ppr}$, and
hence $\Ppa \sqrt{\Ppr} <1$ is the material condition.
Now $\mathcal{G}$ is shown in Fig.~\ref{fig2}(b), and stability analysis is quite similar to that for the case (a). \\
(c) In the case $\Ppa - \gamma > 0$ and $\Ppr - 1/\gamma^2 < 0$, $\mathcal{G}$ is always positive, Fig.~\ref{fig2}(c).
The reduced torque $\mathcal{G}$ has a minimum at $\lambda_2 = [(2(1/\gamma^2 - \Ppr)/(\Ppa - \gamma)]^{1/3}$, and this minimum
decreases by lowering the temperature from $T_2$ to $T_2'$.
Again, if one starts at the point A where $\mathcal{G} > G\s{ext}/\mu A_0 R_1$, the engine will move to operate at the point B. \\
(d) If $\Ppa - \gamma < 0$ and $\Ppr - 1/\gamma^2 > 0$, then $\mathcal{G}<0$; there are no solutions for $G\s{ext} > 0$.  Reversing the
external torque, and cooling rather than heating to $T_2$, reverses the motor and we have an analogy to the case (c).

We adopt a simple freely jointed rod model for the polymer backbones, with a step length $a$ in the isotropic state.
Then the step lengths are $\lz = a(1+2Q) $ and $\lp = a(1-Q)$, with the nematic order parameter $ 0 \leq Q \leq 1$.
Although crude, this model quite accurately describes a wide range of LCEs~\cite{warnerbook:07,finkelmann:01t}, and, in particular, the
development of photoforce~\cite{knezevic:13}.

The material parameters $\Ppa$ and $\Ppr$ now read $\Ppa = (1+2Q_1)/(1+2Q_2)$ and $\Ppr = (1-Q_1)/(1-Q_2)$, where
$Q_1=Q(T_1)$ and $Q_2=Q(T_2)$. Then the above material conditions can be expressed in terms of
$Q_1$ and $Q_2$. For example, the condition $\Ppa > 1/\sqrt{\Ppr}$ of case (a) is $g(Q_1)>g(Q_2)$,
where $g(Q) = 3Q - 4Q^3$. Since $Q_1 > Q_2$, the condition $g(Q_1)>g(Q_2)$ is satisfied whenever $Q_1 \leq 1/2$. Further, for every $Q_1$
lying in the interval $1/2 < Q_1 < \sqrt{3}/2$ one can find a threshold value of $Q_2$ below which the condition $g(Q_1)>g(Q_2)$ holds.
Lastly, for $Q_1 \geq \sqrt{3}/2$ the condition $g(Q_1)>g(Q_2)$ cannot be satisfied.
The material condition $\Ppa < 1/\sqrt{\Ppr}$ of (b) is $g(Q_1)<g(Q_2)$, and corresponding
conditions in terms of $Q$ are easily obtained. In the case (c) one has $\gamma < \min(\Ppa, 1/\sqrt{\Ppr})$.
If the temperature $T_2$ is above the nematic-isotropic transition temperature one has $Q_2 = 0$, and consequently $\Ppa > 1/\sqrt{\Ppr}$ for all $Q_1 < \sqrt{3}/2$.

We estimate the efficiency of the engine as $\eta = P\s{out}/P\s{in}$, where $P\s{in}$ is the power needed
to heat an incoming element of elastomer at temperature $T_1$ to temperature $T_2$, and $P\s{out}$ is the corresponding power output.
The input power can be expressed as $P\s{in} = C\s{p} \Delta T A_0 (d l_0/d t)$, where $dl_0$ is the length of
a piece of elastomer in the formation state, currently stretched by $\lambda_1$.
Here $C\s{p}$ denotes the isobaric heat capacity per unit volume of elastomer and $\Delta T = T_2 - T_1$.
For an element of elastomer lying on the wheel of radius $R_2$ one can write $\lambda_1 (dl_0/dt) = \omega R_2$,
where $\omega$ is the angular velocity. The output power is $P\s{out} = G\s{ext} \omega$.  The reduced efficiency, $\widetilde{\eta}  = \eta C\s{p} \Delta T/\mu $, arises through Eq.~(\ref{reducedg})  and is
\be
 \widetilde{\eta} = (\gamma - 1) \left [ \lambda_2^2 (\Ppa - \gamma) - \left ( \Ppr - \frac{1}{\gamma^2}
\right ) \frac{1}{\lambda_2} \right ],
\label{eta}
\ee
where $\lambda_2$ is given by Eq.~(\ref{l2}).

We roughly estimate  $C\s{p} \Delta T$ using the latent heat per unit volume of an idealized, sharp (first-order) nematic--isotropic transition. Its approximate value is $2 \times 10^6 {\rm Jm}^{-3}$~\cite{warnerbook:07}.
Since the isotropic shear modulus is of the order $\mu \sim 10^5 - 10^6 {\rm Jm}^{-3}$, then $\mu/C\s{p} \Delta T$ can be up to 0.5.
For photoelastomers the energy input $C\s{p} \Delta T$ represents $\varepsilon n\s{dye}$, where $\varepsilon$ is the photon energy and
$n\s{dye}$ is the number density of dye molecules~\cite{knezevic:13a}, giving for $\mu/\varepsilon n\s{dye}$ an estimate of the same order as that
for $\mu/C\s{p} \Delta T$.

As we have seen, when $Q_2 = 0$ the constraint is $\Ppa > 1/\sqrt{\Ppr}$, which restricts us to cases (a) or (c). The efficiency (\ref{eta}) depends on four dimensionless quantities: the order parameter $Q_1$ (through $\Ppa = 1 + 2Q_1$
and $\Ppr = 1 - Q_1$), the reduced lengths $\widetilde{L}$ and $\widetilde{L}_0$ (through $\lambda_2$), and the ratio of the wheel radii $\gamma$.
Clearly, the efficiency increases with increasing order $Q_1$.
Regarding the efficiency as a function of $\widetilde{L}_0$, $\widetilde{\eta}$ takes quite large values for $\widetilde{L}_0 \ll 1$ as well as for $\widetilde{L}_0 \gg 1$. Similarly, the efficiency increases with increasing $\widetilde{L}$.
The engine can operate only if certain physical constraints are satisfied, implying that $\widetilde{L}_0$ and $\widetilde{L}$ are not completely independent of each other.
First, to obtain a contractile force the stretch $\lambda_2$ should be greater than the natural contraction $\lambda\s{m}$ of the freely suspended elastomer, which can be expressed as $\gamma < \widetilde{L}/ (\widetilde{L}_0 \lambda\s{m} - \widetilde{L} - 2)$. This condition, together
with $\gamma>1$, implies that $\widetilde{L}_0 \lambda\s{m}/2 - 1 < \widetilde{L} < \widetilde{L}_0 \lambda\s{m} - 2$.
Besides, since one cannot mechanically contract a thin elastomer below its natural length then $\lambda_1>1$. The ratio $\gamma$ is thus limited from below, $\gamma > (\widetilde{L}_0 - \widetilde{L})/(\widetilde{L}+2)$.
In addition, to avoid slack when the elastomer is stretched from its formation state and wound around the pulleys,
one has $\gamma > \widetilde{L}_0 - 2 \widetilde{L} - 1$. In summary:
\bea
&\gamma& >\max \left ( 1, \frac{\widetilde{L}_0 - \widetilde{L}}{\widetilde{L}+2}, \widetilde{L}_0 - 2 \widetilde{L} - 1 \right ), \nonumber \\
&\gamma& < \min \left ( \Ppa, \frac{\widetilde{L}}{\widetilde{L}_0 \lambda\s{m} - \widetilde{L} - 2} \right ),
\label{condgamma}
\eea
taking into account that $\gamma<\Ppa$  ---  a consequence of the reasonable assumption
$Q_1 < \sqrt{3}/2$.
\begin{figure}
  \includegraphics[width=8.4cm]{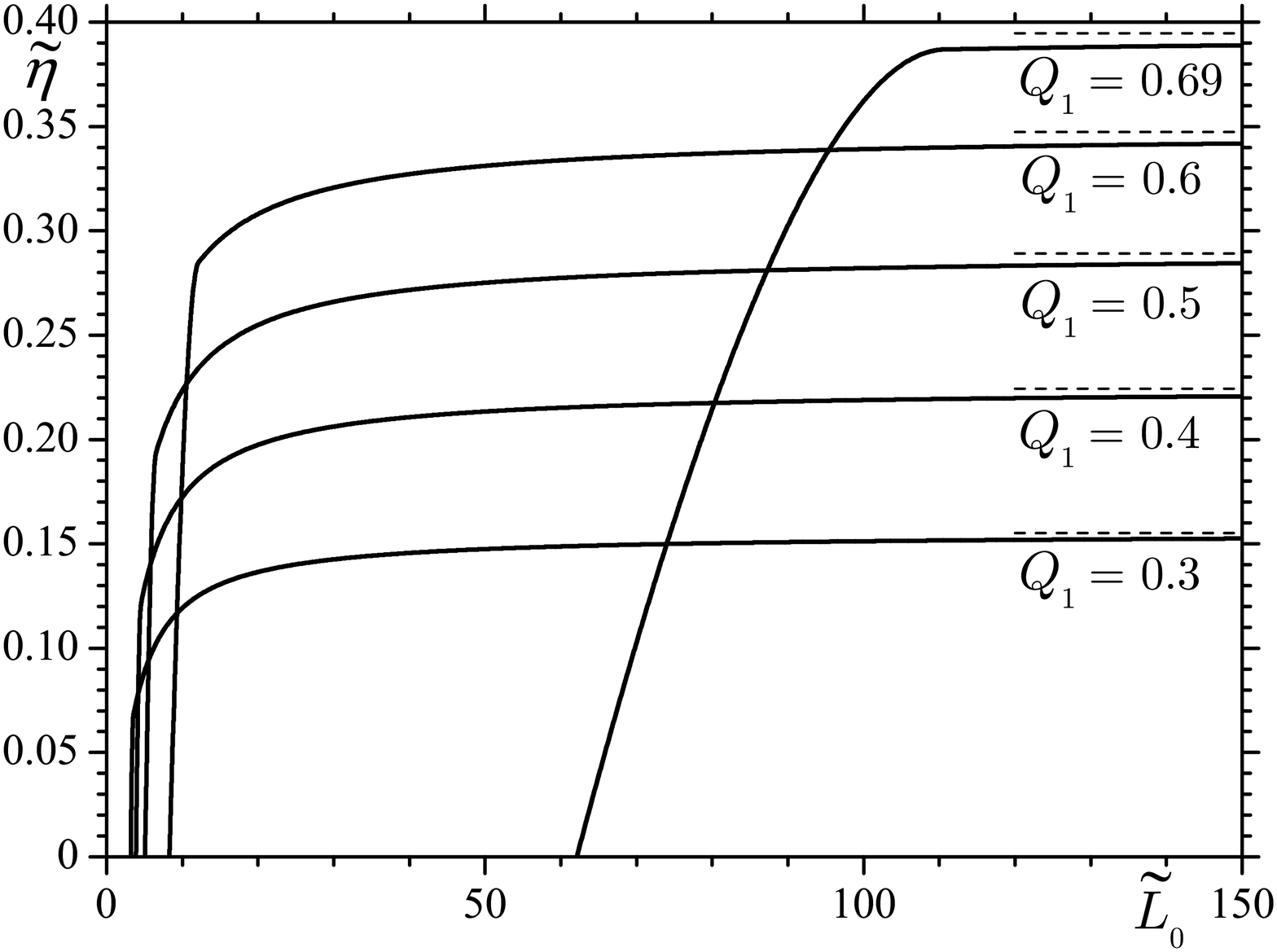}
  \caption{The reduced efficiency $\widetilde{\eta}$ as a function of
  $\widetilde{L}_0$ for different values of $Q_1$ (solid lines). Along the plateau of each
  of these curves the optimal values of $\gamma$ change only slightly, taking the values
  $\gamma \approx 1.22, 1.28, 1.34, 1.40, 1.47$ for $Q_1 = 0.3, 0.4, 0.5, 0.6, 0.69$, respectively.
  The dashed lines correspond to  $\widetilde{L}_0 \gg 1$. The efficiency $\eta = \widetilde{\eta}  \mu/(\varepsilon n\s{dye})$ can go up to 20\%.
  }
\label{fig3}
\end{figure}

The optimal value of $\widetilde{\eta}$ is obtained by choosing $\widetilde{L}$ as large as possible, $\widetilde{L} = \widetilde{L}_0 \lambda\s{m} - 2$, then maximizing $\widetilde{\eta}$ with respect to $\gamma$ and making sure that the constraints (\ref{condgamma}) are satisfied.
Numerical results for the reduced efficiency $\widetilde{\eta}$ are shown in Fig.~\ref{fig3}.
Optimal values of $\widetilde{\eta}$ are reached already for moderate  $\widetilde{L}_0 \approx 20$
for $Q_1 \lesssim 0.6$. The efficiency $\eta = \widetilde{\eta} \mu/(\varepsilon n\s{dye})$ can go up to 20\% for $Q_1 \approx 0.7$ in the optical case. For $Q_1 > 0.7$, the no slack condition (\ref{condgamma}) is violated.
Such high values of $Q_1$ in Fig.~\ref{fig3} are perhaps unphysical in side-chain LCEs, but they represent the high anisotropy in $\Ppa$ and $\Ppr$ found in main-chain elastomers serving as working materials. Their $\Ppa$ and $\Ppr$ values are more
extreme, even at normal values of $Q_1$, and $1/\lambda\s{m}$ can be as large as 350\%~\cite{tajbakhsh:01}.  For photo-engines, the thickness of the elastomer band depends critically on the light intensity. Non-linear absorption processes determine optical penetration and force dynamics~\cite{knezevic:13,knezevic:13a}; for mm thicknesses intensities of  $10 - 100 \textrm{mW/cm}^2$ are required -- smaller than maximal insolation.

In summary the thermo-optical contraction of nematic elastomers can be used to harness thermal or optical
energy to generate mechanical energy. Further efficiency can be gained in both material design and geometric improvements to
the engine.

M. K. acknowledges support from the Winton Programme for the Physics of Sustainability and the Cambridge Overseas Trust, and M. W. thanks
the Engineering and Physical Sciences Research Council (UK) for a Senior Fellowship. We are grateful to E. M. Terentjev
and P. Palffy-Muhoray for useful discussions.

\end{document}